# Significant crosstalk reduction using all-dielectric CMOS-compatible metamaterials

Amin Khavasi, Lukas Chrostowski, Zeqin Lu and Richard Bojko

*Abstract*— **A recent computational result suggests that highly confined modes can be realized by all-dielectric metamaterials (S. Jahani et. al., Optica 1, 96 (2014)). This substantially decreases crosstalk between dielectric waveguides, paving the way for high-density photonic circuits. Here, we experimentally demonstrate, on a standard silicon-on-insulator (SOI) platform, that using a simple metamaterial between two silicon strip waveguides results in about a 10-fold increase in coupling length. The proposed structure may lead to significant reduction in the size of devices in silicon photonics.**

*Index Terms*— **Silicon photonics, Metamaterials, Crosstalk**

## I. Introduction

Silicon photonics has attracted significant attention for realizing nanophotonic devices during the past two decades due to its compatibility with the well-established complementary metal oxide semiconductor (CMOS) technology [1]. However, the diffraction limit of light is a fundamental barrier to achieve more miniaturized photonic circuits using dielectric materials [2]. Plasmonics goes beyond the diffraction limit making possible the realization of subwavelength optical devices [3]–[6]. However, the high ohmic losses in conventional plasmonic materials hamper their application as waveguides [2].

It is possible to manipulate the flow of light inside dielectric waveguides by locally engineering the dielectric permittivity at the subwavelength scale [7]. These structures, namely all-dielectric metamaterials have been used for several applications such as waveguide loss reduction [8], achieving giant birefringence [9], efficient couplers and multiplexers [10] and polarization beam splitters with relaxed fabrication tolerance [11]. More recently, a new approach has been proposed to confine light below the diffraction limit inside a dielectric waveguide using all-dielectric metamaterials [12], [13]. The strategy relies on controlling the optical momentum of evanescent waves using all-dielectric metamaterials. To this end, a relaxed version of total internal reflection (TIR), an essential condition in dielectric waveguides, is introduced [12]. It is demonstrated that if the perpendicular (with respect to the interface) component of the dielectric tensor of the cladding can be smaller than the dielectric constant of the core, the TIR condition will be satisfied. On the other hand, the parallel component of the dielectric tensor of the cladding should be as high as possible to reduce the penetration depth [12]. This extreme anisotropy can be realized by artificially structured media using available lossless dielectrics.

A multilayer structure consisting of two materials with high index contrast and layer thicknesses far below the wavelength of light can be used to realize such an anisotropic medium. This approach has been numerically tested and it has been proven to be successful in reducing crosstalk between one-dimensional (1D) slab waveguides [12]. For two-dimensional (2D) strip waveguides, however, this approach should be modified.

In this letter, we propose a similar structure made of silicon ribbons to reduce crosstalk between strip waveguides. The proposed structure is fabricated on a standard SOI platform. Experimental results show that the crosstalk can be significantly reduced using the proposed structure. Two strip waveguides with a center-to-center distance of just 1 μm are tested at telecommunications wavelength of 1.55 μm. Our experimental results show that the coupling length (the distance over which the input power is completely coupled to the adjacent waveguide) is increased from ~0.393 mm to ~3.88 mm.

## II. The proposed structure

The cross section diagrams of the two strip waveguides, which are studied in this work, are shown in Figure 1. Each strip waveguide has a rectangular cross section made of silicon, surrounded by $SiO_2$. The height of the waveguides is 220 nm, which is the thickness of the silicon layer in the SOI wafers commonly used in silicon photonics. We choose the width of waveguides to be 500 nm to achieve high confinement and to maintain nearly single quasi-transverse-electric (TE) mode operation [1]. The waveguides have one propagating TE-like mode and only this mode will be considered throughout this work.

The waveguides' separation is 1000 nm so their coupling is not negligible. Two kinds of coupling are investigated: (1) conventional coupling through $SiO_2$ (Figure 1(a)) and (2) coupling through a metamaterial (Figure 1(b)). The metamaterial consists of Si and $SiO_2$ layers (in $x$ direction) with widths of $w$ and $a$. The height of the Si layer is 220 nm according to the SOI wafer used in the fabrication.

A. Khavasi is with Electrical Engineering Department, Sharif University of Technology, Tehran 11155-4363, Iran (e-mail: khavasi@sharif.edu).

L. Chrostowski and Zeqin Lu are with the Department of Electrical and Computer Engineering, University of British Columbia, Vancouver, BC V6T 1Z4, Canada (e-mail: lukasc@ece.ubc.ca).

Richard Bojko is with the Department of Electrical Engineering, University of Washington, Campus Box 352500, Seattle, Washington 98195, USA

If we inject the light to one of the waveguides, after some distance, say $L$, some of the power will be coupled to the second waveguide. The power in the first (direct) and the second (coupled) waveguides (normalized to the input power) are given by

$$P_D(L) = \cos^2(\Delta\beta L), \quad (1)$$

and

$$P_C(L) = \sin^2(\Delta\beta L), \quad (2)$$

respectively. In these equations $\Delta\beta = |\beta_1 - \beta_2|/2$ where $\beta_1$ and $\beta_2$ are the propagation constants of the super-modes of the coupled waveguide system [14]. Considering a sinusoidal form for $P_D$ as given in (1), the coupling length, the distance over which the transmitted power to the direct port becomes zero, is $L_C = \pi / 2\Delta\beta$ or

$$L_C = \frac{\pi L}{2\cos^{-1}\left(\sqrt{P_D}\right)} \quad (3)$$

The propagation constants of the modes are numerically calculated using Lumerical MODE Solutions. The modal fields (the major component of the electric field $E_x$) are also calculated for the super-modes and they are illustrated in Figure 1(c) and (d) for the conventional and metamaterial (with $w$ = 80 nm) structures, respectively. The effective indexes of super-modes for both structures are also plotted in Figure 2. It is obvious that the difference between the effective indexes is decreased by using metamaterial.

The simulation results of $L_C$ in terms of $w$ are plotted in Figure 3. These results show that $w$ = 60–80 nm leads to minimum coupling. We set $w$ = 80 nm (and thus $a$ = 113 nm) which is sufficiently larger than the 60 nm resolution of the lithography system used for the fabrication of the device. A scanning electron microscope (SEM) picture of the fabricated device with $w$ = 80 nm and $a$ = 113 nm, is illustrated in Figure 4(a).

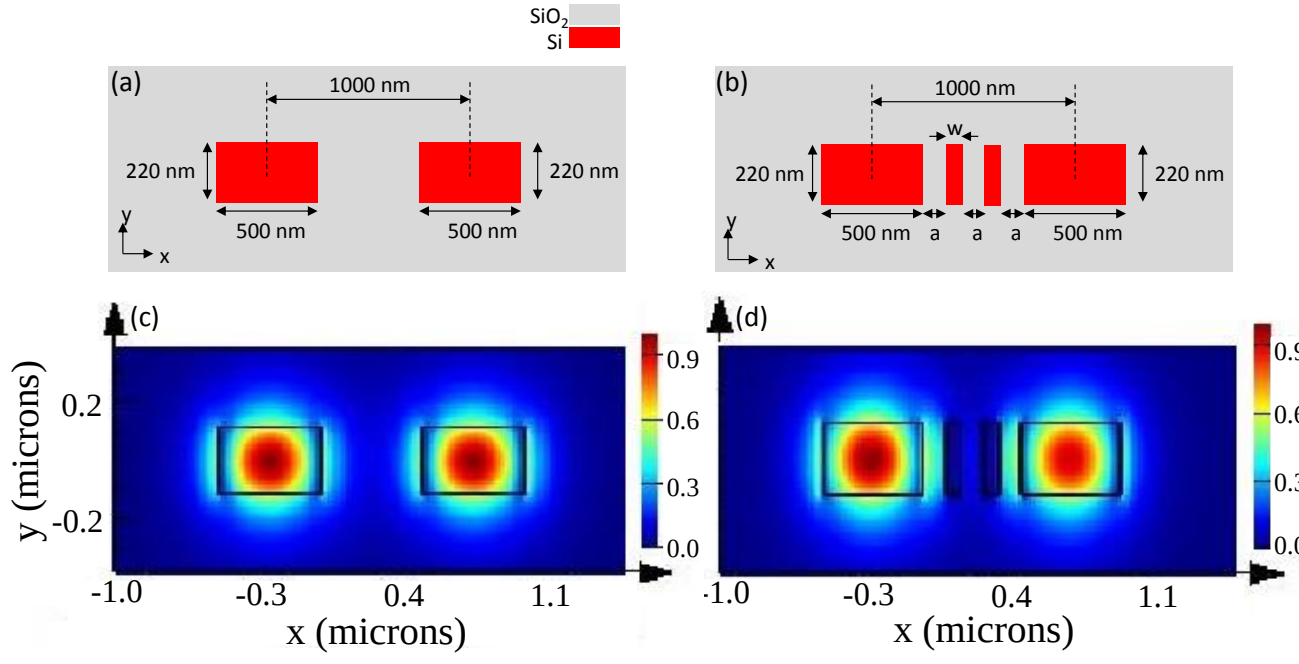

Figure 1- Cross section of two strip Si waveguides coupled to each other through (a) $SiO_2$ and (b) metamaterial and their corresponding modal fields are shown in (c) and (d), respectively.

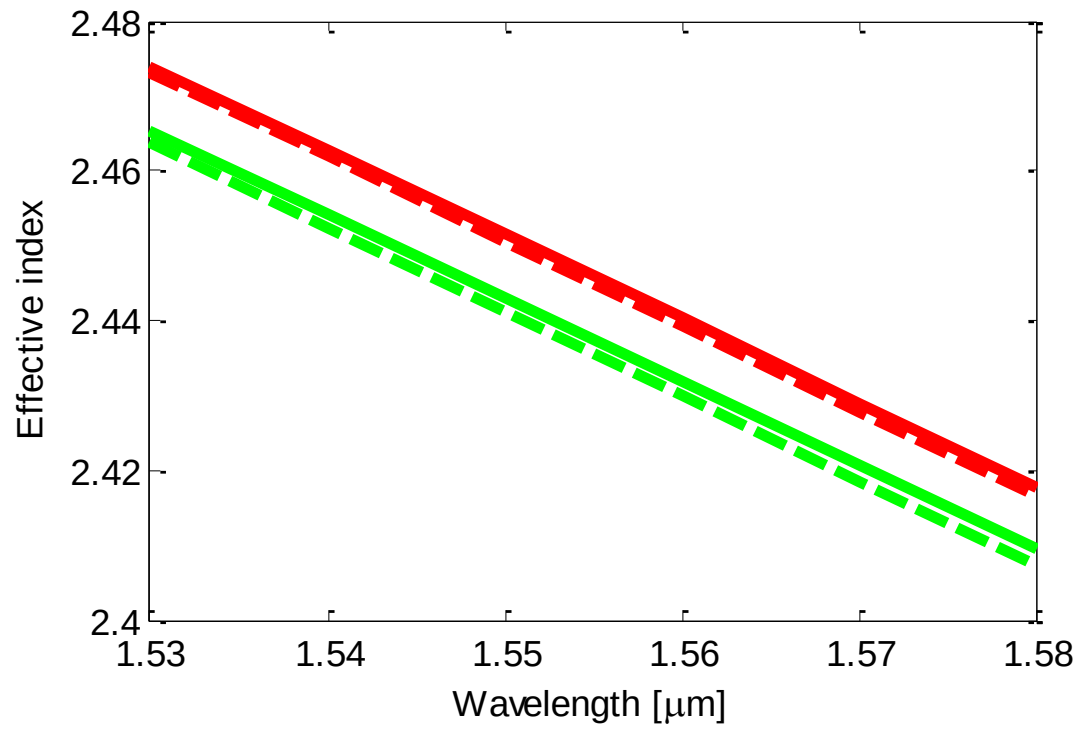

Figure 2- The effective index of the super-modes of the coupled waveguides with conventional (green) and metamaterial (red) couplings as a function of wavelength. For the metamaterial case $w$ has been set to 80 nm.

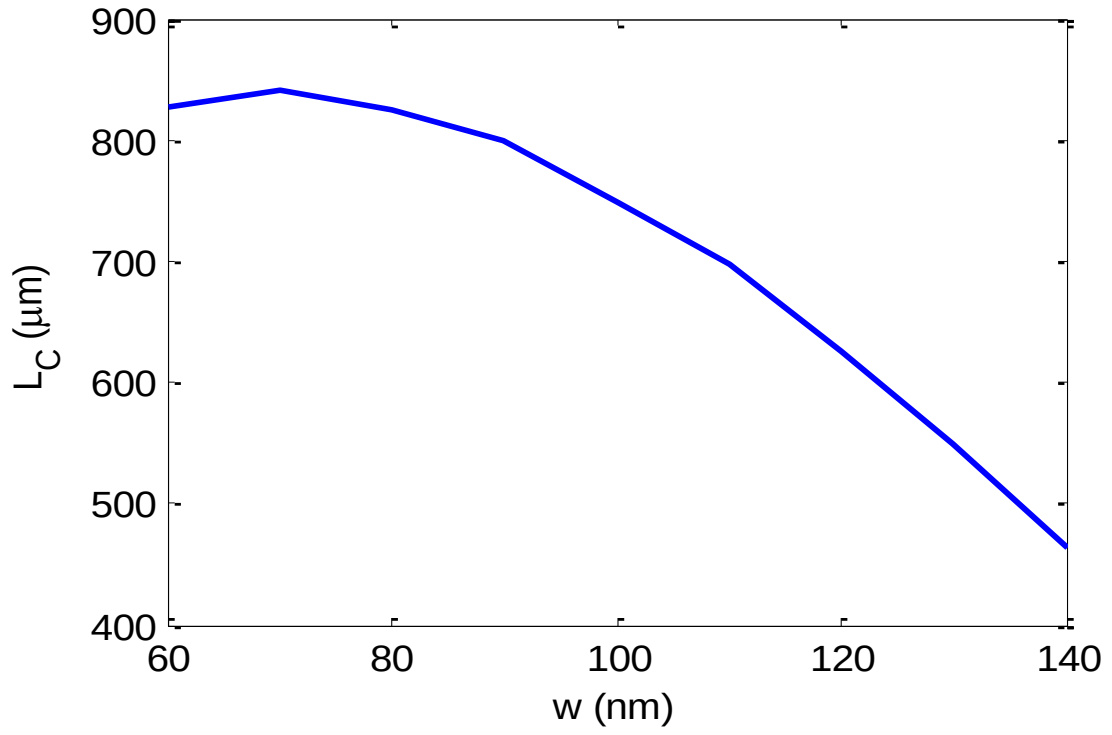

Figure 3- The coupling length versus the width of the Si layer of metamaterial ($w$).

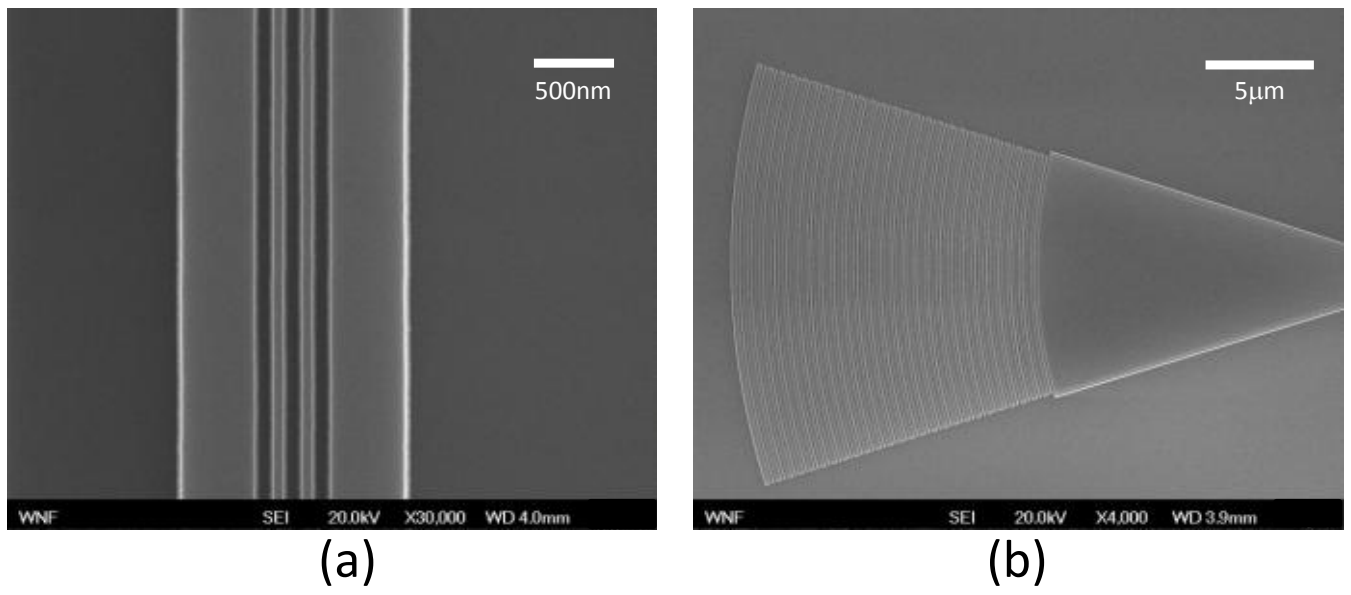

Figure 4- SEM images of (a) the fabricated waveguides with metamaterial between them for coupling reduction and (b) the grating coupler used for coupling light into and out of the structure.

## III. Fabrication and Measurement Setup

The devices were fabricated using 100 keV Electron Beam Lithography [15]. The fabrication used SOI wafers with 220 nm thick silicon on 3 μm thick silicon dioxide. The substrates were 25 mm squares diced from 150 mm wafers. After a solvent rinse and hot-plate dehydration bake, hydrogen silsesquioxane resist (HSQ, Dow-Corning XP-1541-006) was spin-coated at 4000 rpm, then hotplate baked at 80 °C for 4 minutes. Electron beam lithography was performed using a JEOL JBX-6300FS system operated at 100 keV energy, 8 nA beam current, and 500 $\mu$m exposure field size. The machine grid used for shape placement was 1 nm, while the beam stepping grid, the spacing between dwell points during the shape writing, was 6 nm. An exposure dose of 2800 $\mu C/cm^2$ was used. The resist was developed by immersion in 25% tetramethylammonium hydroxide for 4 minutes, followed by a flowing deionized water rinse for 60 s, an isopropanol rinse for 10 s, and then blown dry with nitrogen. The silicon was removed from unexposed areas using inductively coupled plasma (ICP) etching in an Oxford Plasmalab System 100, with a chlorine gas flow of 20 sccm, pressure of 12 mT, ICP power of 800 W, bias power of 40 W, and a platen temperature of 20 °C, resulting in a bias voltage of 185 V. During etching, chips were mounted on a 100 mm silicon carrier wafer using perfluoropolyether vacuum oil.

To characterize the devices, a custom-built automated test setup [1] with automated control software written in Python was used. An Agilent 81600B tunable laser was used as the input source and Agilent 81635A optical power sensors as the output detectors. The wavelength was swept from 1500 to 1600 nm in 10 pm steps. A polarization maintaining (PM) fiber was used to maintain the polarization state of the light, to couple the TE polarization into/out of grating couplers [16]. The SEM image of one of the fabricated grating couplers is depicted in Figure 4(b).

In order to estimate the power transmitted to the direct and coupled ports, we use the schematic depicted in Figure 5. In this circuit the power is injected to one of the waveguides through the input grating coupler and the two waveguides are coupled to each other over a distance *L*. Then, the two waveguides are separated and each experiences different path lengths. Finally, the two waveguides are combined by a Y-branch and connected to the output grating coupler.

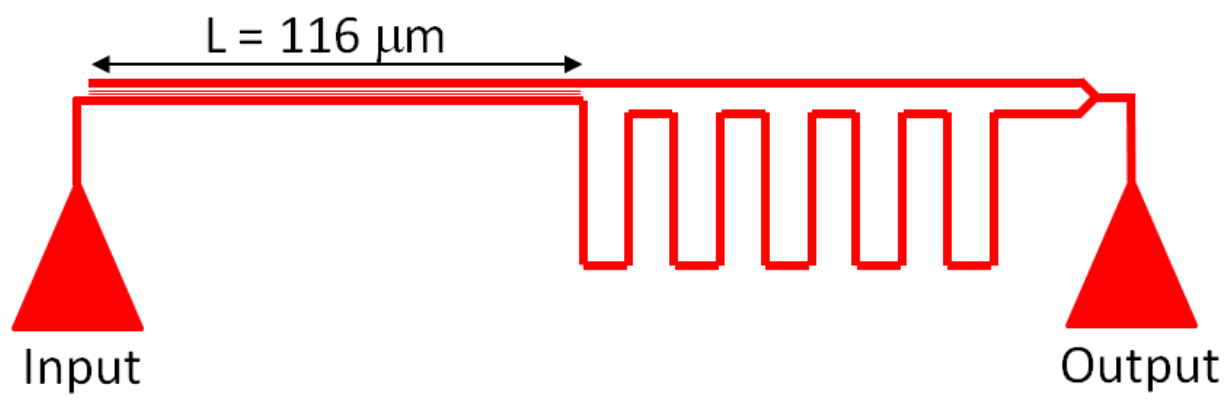

Figure 5- The schematic of the fabricated circuit for measurement of power transmitted to the direct and coupled ports.

## IV. Results and Discussion

The power measured at the output grating coupler for two strip waveguides with the conventional coupling and the metamaterial coupling are plotted in Figure 6 (solid blue curves). Maxima and minima are clearly seen in these figures, which are, respectively, due to constructive and destructive interference between the direct and coupled waves.

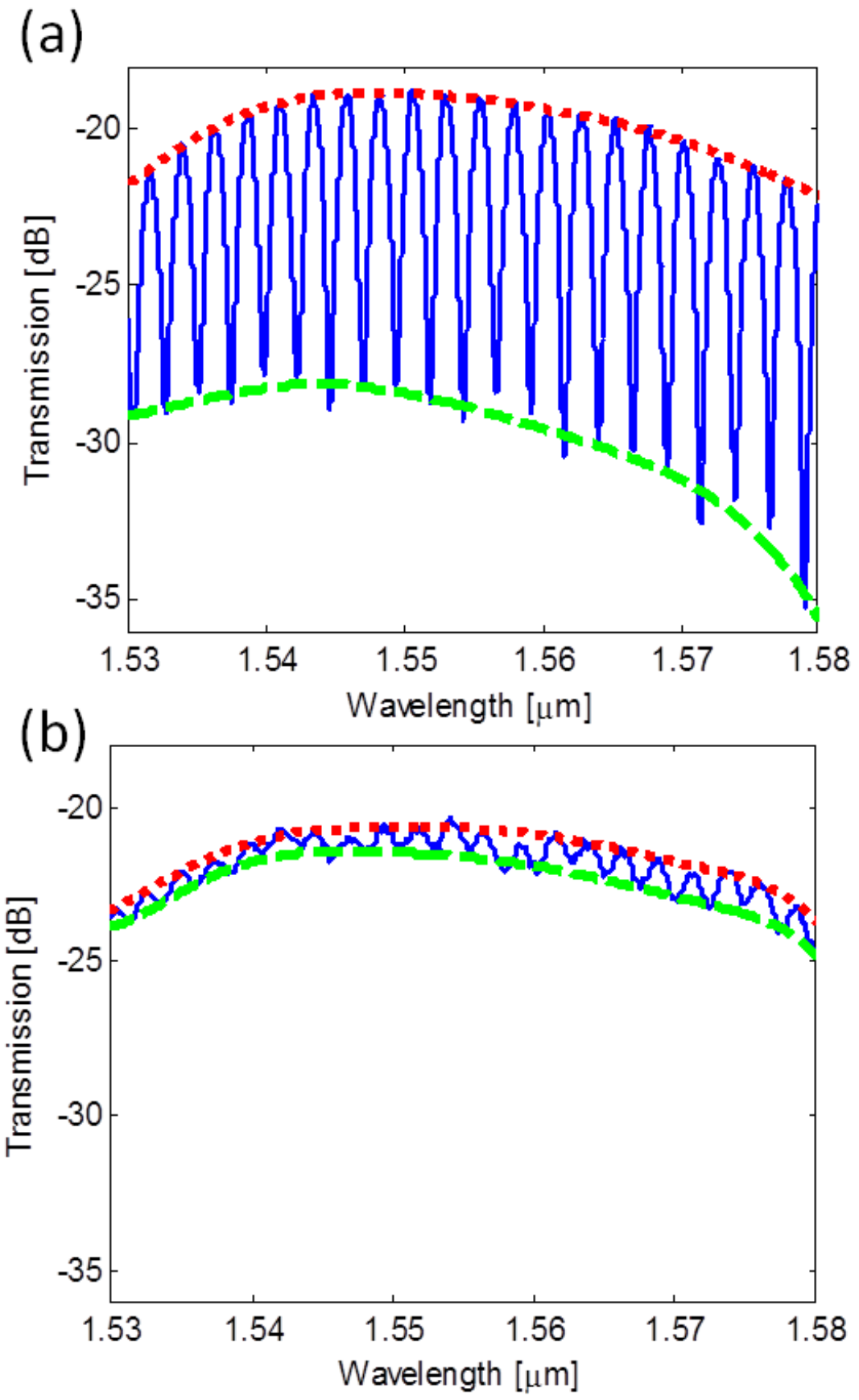

Figure 6- The power spectra measured at the output grating coupler (solid blue curves) for (a) the conventionally coupled waveguides and (b) the waveguides coupled through the metamaterial. Polynomial fits to the maxima (red dotted curves) and the minima of (green dashed curves) the two spectra are also illustrated.

Assuming an ideal Y-branch, the output power for a constructive interference is [1]:

$$P_{\max} \propto |E_{\max}|^2 = \left|\frac{E_D + E_C}{\sqrt{2}}\right|^2 \qquad (4)$$

where $E_D$ and $E_C$ are the electric fields of the waves at the direct and coupled ports, respectively. On the other hand, the output power for a destructive interference is [1]:

$$P_{\min} \propto |E_{\min}|^2 = \left|\frac{E_D - E_C}{\sqrt{2}}\right|^2 \qquad (5)$$

We assume that $E_D > E_C$ and thus according to equations (1) and (2) the distance $L$ should be shorter than the half of the coupling length for all wavelengths (here we choose $L$ = 116 μm). With this assumption and after some straightforward mathematical manipulations, the following relation is obtained for the normalized power transmitted through the direct port:

$$P_D = \frac{\left(\sqrt{P_{\max}/2} + \sqrt{P_{\min}/2}\right)^2}{P_{\max} + P_{\min}} \qquad (6)$$

The normalized power transmitting through the coupled port is simply given as $P_C = 1 - P_D$. It should be noted that, in these calculations, the loss in the waveguides and bends has been neglected.

We use polynomial fits to find $P_{max}$ and $P_{min}$ from the experimental data. These polynomial fits are plotted by red dotted curves and green dashed curves in Figure 6, which are corresponding to $P_{max}$ and $P_{min}$, respectively. Finally, the power transmitted to the direct port is calculated by using (6) and then the coupling length is obtained from (3). The results are shown in Figure 7(a) and (b) for the conventional coupling and the proposed metamaterial coupling, respectively. It is obvious that the coupling strength is substantially reduced by using the proposed metamaterial between the waveguides. More interestingly the experimental results for the proposed metamaterial structure are much better than simulations. We have no justification for these surprising results.

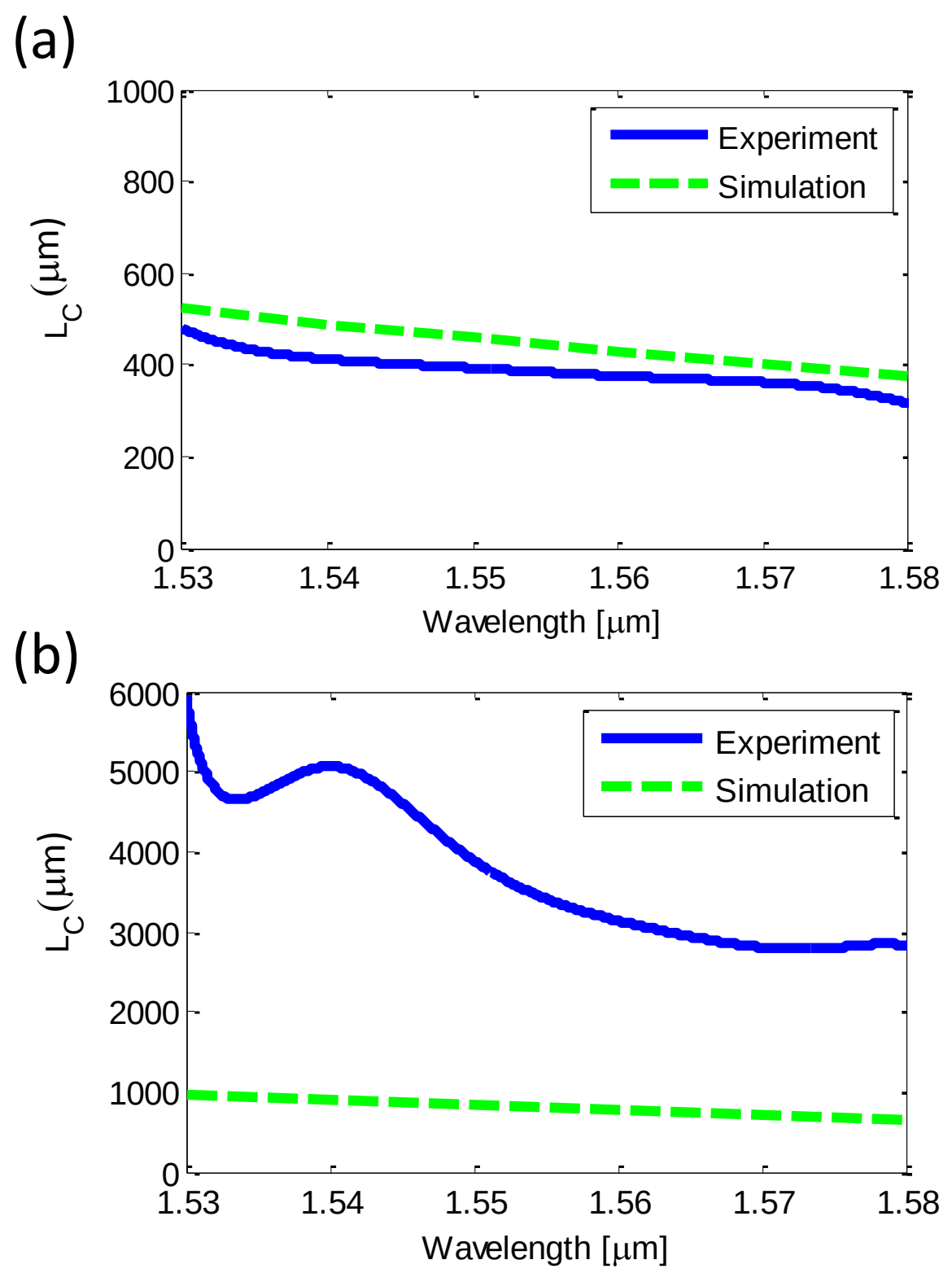

Figure 7- The coupling length for (a) the conventionally coupled waveguides and (b) the waveguides coupled through the metamaterial.

Let us compare the coupling length for the two cases. The coupling lengths at the telecommunications wavelength of 1.55 µm (computed by experimental results) are $L_C$ = 3.88 mm and $L_C$ = 0.393 mm for the proposed structure and the conventional coupling, respectively. Such a large coupling length ($L_C$ = 3.88 mm) allows the waveguides to be fabricated very close to each other, paving the way for ultrahigh-density photonic circuits.

## V. Conclusion

In summary, we have proposed an all-dielectric CMOS-compatible metamaterial for crosstalk reduction between adjacent strip waveguides. The proposed structure has a measured coupling length of about 3.88 mm for 1 µm center-to-center separation between the waveguides at the telecommunications wavelength of 1.55 µm, demonstrating about 10-fold improvement over conventionally coupled waveguides with the same center-to-center separation. This work can lead to significant miniaturization of silicon photonic circuits.

## Acknowledgments:

Fabrication support was provided via the Natural Sciences and Engineering Research Council of Canada (NSERC) Silicon Electronic-Photonic Integrated Circuits (SiEPIC) Program. The devices were fabricated by Richard Bojko at the University of Washington Nanofabrication Facility, part of the National Science Foundation's National Nanotechnology Infrastructure Network (NNIN). Zeqin Lu performed the measurements at The University of British Columbia. We acknowledge Lumerical Solutions, Inc., Mathworks, and KLayout for the design software.